\documentclass[11pt]{article}
\usepackage[colorlinks=true,citecolor=blue,linkcolor=blue]{hyperref}
\usepackage{multirow}
\usepackage[left=2cm,right=2cm,top=2cm,bottom=2cm]{geometry}
\usepackage{cite}
\usepackage{diagbox}
\usepackage{psfrag}
\usepackage{graphicx}
\usepackage{epsfig}
\usepackage{verbatim}
\usepackage{booktabs}
\usepackage{amsmath,amsfonts,amssymb}
\usepackage{pstcol,pst-fill,pst-grad}
\usepackage{pstricks,pst-fill,pst-grad}
\usepackage{euscript}
\usepackage{pstricks}
\usepackage{tabularx}
\usepackage{wrapfig}
\usepackage{slashed}
\usepackage[toc,page]{appendix}
\usepackage{here}
\usepackage{multirow}
\usepackage{ulem}

\date{}
\begin{document}
		\title{\vspace{-3cm}
			\hfill\parbox{4cm}{\normalsize \emph{}}\\
			\vspace{1cm}
			{Insight into the effect of rare earth $RE=Ce,\; Nd,\; Gd$    elements on the physical properties of  Pb-Based Pervoskite Oxides ($PbREO_3$)
		}}
		\vspace{2cm} 
		
		\author{M Agouri$^{1}$, A Waqdim$^{1}$, A Abbassi$^{1,}$\thanks{Corresponding author, E-mail: abbassi.abder@gmail.com}, S Taj$^{1}$, B Manaut$^{1,}$\thanks{Corresponding author, E-mail: b.manaut@usms.ma}, M Driouich$^1$ \\
			{\it {\small$^1$ Laboratory of Research in Physics and Engineering Sciences,}}\\
			{\it{\small Sultan Moulay Slimane University, Polydisciplinary Faculty, Beni Mellal, 23000, Morocco.}}\\
		}	
		\maketitle \setcounter{page}{1}
		\date{\today}
		
		\begin{abstract}
The electronic, elastic, structural, thermoelectric and optical parameters of the Pb-based cubic perovskites $PbREO_3\; (RE=Ce,\; Gd,\; Nd)$ were calculated using the WIEN2k Package. The mechanical optimization and structural stabilities were evaluated. The found results are in full agreement with the experimental ones. The obtained elastic results show that these perovskites behave as stable ductile materials, and their higher $Y$ value gives them a rigid structures. The $PbCeO_3$ presents n-type semiconductor while 
$PbREO_3\; (RE= Gd,\; Nd)$ exhibit p-type semi-conductor with gap energies of $2.60\;eV$, $1.02\;eV$ and $1.55\;eV$, respectively. The optical properties show that $PbREO_3\; (RE=Ce,\; Gd,\; Nd)$ have significant absorption in the visible light region with low reflectivity. The thermoelectric properties present a desired electrical performance at room temperature for all materials. For these generation of materials the effect of RE on Pb-based cubic perovskite oxides can be exploited in manufacturing of thermoelectric, mechanic and optoelectronic devices.			
		\end{abstract}
		
		\maketitle
		\newpage
	\section{Introduction}
The Perovskite oxides are considered as a particular class of materials combining several exclusive properties such as high optical transmittance and high electrical conductivity. They can be multimeric materials with different aspects. They have been considered as a promising material candidates for a wide range of applications such as optoelectronic and spintronic, lasers and UV light-emitting diodes (LED) \cite{1}. Perovskite oxides are characterized by their variable formula, flexibility of structure, numerous properties and broad applications. Diverse experimental and theoretical studies have shown perovskite oxides with various properties, such as magnetic \cite{2}, photoelectric \cite{3} and luminescent \cite{4}...etc.\\
Pb-based perovskite oxides have many interest to researchers in the last decade. They have found important applications in a variety of devices and industrial applications such as sensors, storage devices and are very favourable for advanced technological applications \cite{5, 6}. They have been investigated not only for their flexible phases but also for their different interesting properties.\\
Rare-earth (RE) elements have excellent properties \cite{4, 7}, especially magnetic, luminescent, optoelectronic and electrical properties that make them valuable for industrial applications and manufacturing products. These properties are provided in particular from the electrons located in the $4f$ orbitals \cite{7}.\\
RE ions are widely utilized as dopants to improve the different physical and chemical properties. They have been extensively included into perovskite oxides to improve and develop their performances in novel technologies. Therefore, the approach of incorporating Rare-earth ions into perovskite oxides opens new fields in developing multi-functional RE-containing perovskite compounds \cite{8}.\\
Perovskite-oxides have an ideal cubic structure. However, they often undergo one or more structural phase transitions. At room temperature with a pressure of $6.4\;GPa$, the $PbVO_3$ presents a cubic phase, whereas at ambient pressure it exhibit a tetragonal structure \cite{9}. The La-based perovskite oxide $LaCoO_3$ presents a rhombohedral distortion \cite{10}. Whereas, in the case of the $LaMnO_3$ compound, a cubic deformation is observed \cite{11}.\\
Among Pb-based perovskite oxides, $PbTiO_3$ is mainly studied due to its simplicity synthesizes. K. Shahzad et al \cite{12} prepared and determined its crystal structure using Neutron and X-ray Diffractions, respectively. In other theoretical works, the ferroelectric, structural, electronic and magnetic properties of $PbBO_3\; (B=Cr,\; Co,\; Cu,\; Mn\; and \;V)$ were investigated \cite{13}. A. Kadiri et al \cite{14} calculate the magnetic properties of tetragonal $PbVO_3$ by the Monte Carlo simulation method. The optical, dynamic and thermodynamic properties of cubic $BaCeO_3$ were investigated, and they showed that this compound have a semiconductor behavior with a direct band gap ($2.17\; eV$) \cite{15}.\\

The properties observed for Pb-based perovskite oxide materials can be suggested to aim their application for the latest technology. Therefore, we suggest three different materials obtained from the insertion of rare-earth elements $Ce$, $Nd$ and $Gd$ that are selected due to their improved chemical and physical properties. These compounds are not widely treated in the literature. Our study will based on the Density Functional Theory (DFT) using FP-LAPW (Full Potential Linearized Augmented Plane Wave) to study the elastic, electronic, structural, optical and thermoelectric properties of cubic $PbREO_3\; (RE=Ce,\; Nd\; and \;Gd)$ perovskite oxides using GGA and new modified Becke-johnson (nmBJ).\\
The organization of the recent paper can be presented into three parts. Section 2 treats the computational procedures. Section 3 deals with discussion of the obtained results about the physical properties studied, followed by a conclusion in the last section.
\section{Computational details}
The objective of this paper is the investigation of structural, elastic, electronic, optical and thermoelectric properties of cubic $PbREO_3\; (RE=Ce,\; Nd\; and \;Gd)$ perovskite oxides. 
We used the FP-LAPW (Full Potential Linearised Augmented Plane Wave) method within the density functional theory (DFT) introduced in the WIEN2k code \cite{16}. For structural and elastic properties, we applied the generalized gradient approximation presented by Perdew-Burke-Ernzehol (GGA-PBE) \cite{17}. The optimized crystalline structure is obtained by Birch-Murnaghan fits \cite{18}. Elastic properties have been investigated using ElaStic-1.1 package \cite{19}.\\
To investigate other properties (electronic, optical and thermoelectric), we used the new modified Becke-Johnson exchange potential (nmBJ-GGA) \cite{20}. This approximation was proposed by Koller, Tran and Blaha to re-parametrize the parameter $c$ into the modified Becke-Johnson (mBJ) which is given by:
\begin{equation}
	\vartheta^{nmBJ}_{x}(r)=c\vartheta^{BR}_{x}(r)
	+(3c-2)\dfrac{1}{\pi}\sqrt{\dfrac{5}{12}}
	\sqrt{\dfrac{2E(r)}{\rho(r)}},
\end{equation}
where $E(r)$ is the Kohn-Sham kinetic energy density and $\rho(r)$ is the electron density. The parameter $\vartheta^{BR}_{x}(r)$ presents the Becke-Roussel (BR) exchange potential. The new parameterization of c was suggested from the density, and it reads:
\begin{equation}
	c= \mu + \nu \sqrt{\dfrac{1}{V_{cell}}\int_{cell}\dfrac{1}{2}\left( \dfrac{\mid\nabla\rho^{\uparrow}(r')\mid}{\rho_{\uparrow}} +\dfrac{\mid\nabla\rho^{\downarrow}(r')\mid}{\rho_{\downarrow}}\right) dr^{'3}}
\end{equation}
The $\mu$ and $\nu$ have been chosen according to the experimental band gap fits.\\
The separation energy between core and valence electrons is $-9.0\; Ry$. The numbers of plane waves are limited by $R_{MT} \times K_{max} = 8$. The $l_{max}$ parameter was taken to be 10 and the Fourier expanded change density is $G_{max} = 16$. The integration of first Brillouin zone is realized with ($12 \times 12 \times 12$) k-points grid in reciprocal space. The corresponding $R_{MT}$ values for each atom is presented in table \ref{Table:1}. Thermoelectric properties for the $PbREO_3\; (RE=Ce,\; Gd\; and \;Nd)$ have also been investigated using BoltzTraP package \cite{21}.
\begin{table}[H]
	\begin{center}
		\begin{tabular}{ |p{2cm}|p{2cm}|p{2cm}|p{2cm}|  }
			\hline
			\multicolumn{4}{|c|}{\textbf{PbCeO$_3$}}  \\
			\hline
			Atoms     & Pb   &  Ce  & O\\
			\hline
			$R_{MT}$  & 2.50 & 2.23 & 1.92\\
			\hline
			\multicolumn{4}{|c|}{\textbf{PbNdO$_3$}} \\
			\hline
			Atoms     & Pb   &  Nd  & O\\
			\hline
			$R_{MT}$  & 2.50 & 2.25 & 1.84\\
			\hline
			\multicolumn{4}{|c|}{\textbf{PbGdO$_3$}} \\
			\hline
			Atoms     & Pb   &  Gd  & O\\
			\hline
			$R_{MT}$  & 2.50 & 2.22 & 1.81\\
			\hline
		\end{tabular}
		\caption {$R_{MT}$ values of $PbREO_3\; (RE=Ce,\; Nd\; and \;Gd)$ compounds.}
		\label{Table:1}
	\end{center}
\end{table}
\section{Results and discussions}
\subsection{Structural properties}
\begin{figure}[H]
	\centering
	\includegraphics[scale=0.4]{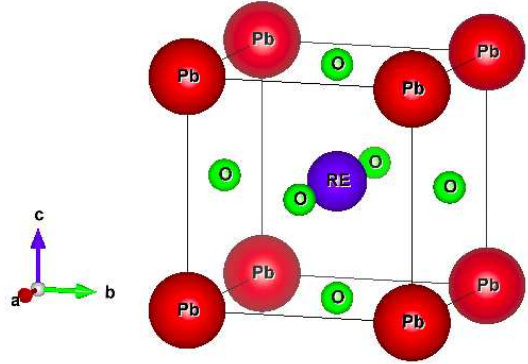}
	\caption{Perovskite cubic  structure of $PbREO_3\; (RE=Ce,\; Nd\; and \;Gd)$ \cite{22}.}
	\label{Figure:1}
\end{figure}

\begin{figure}[H]
	\centering
	\includegraphics[scale=0.31]{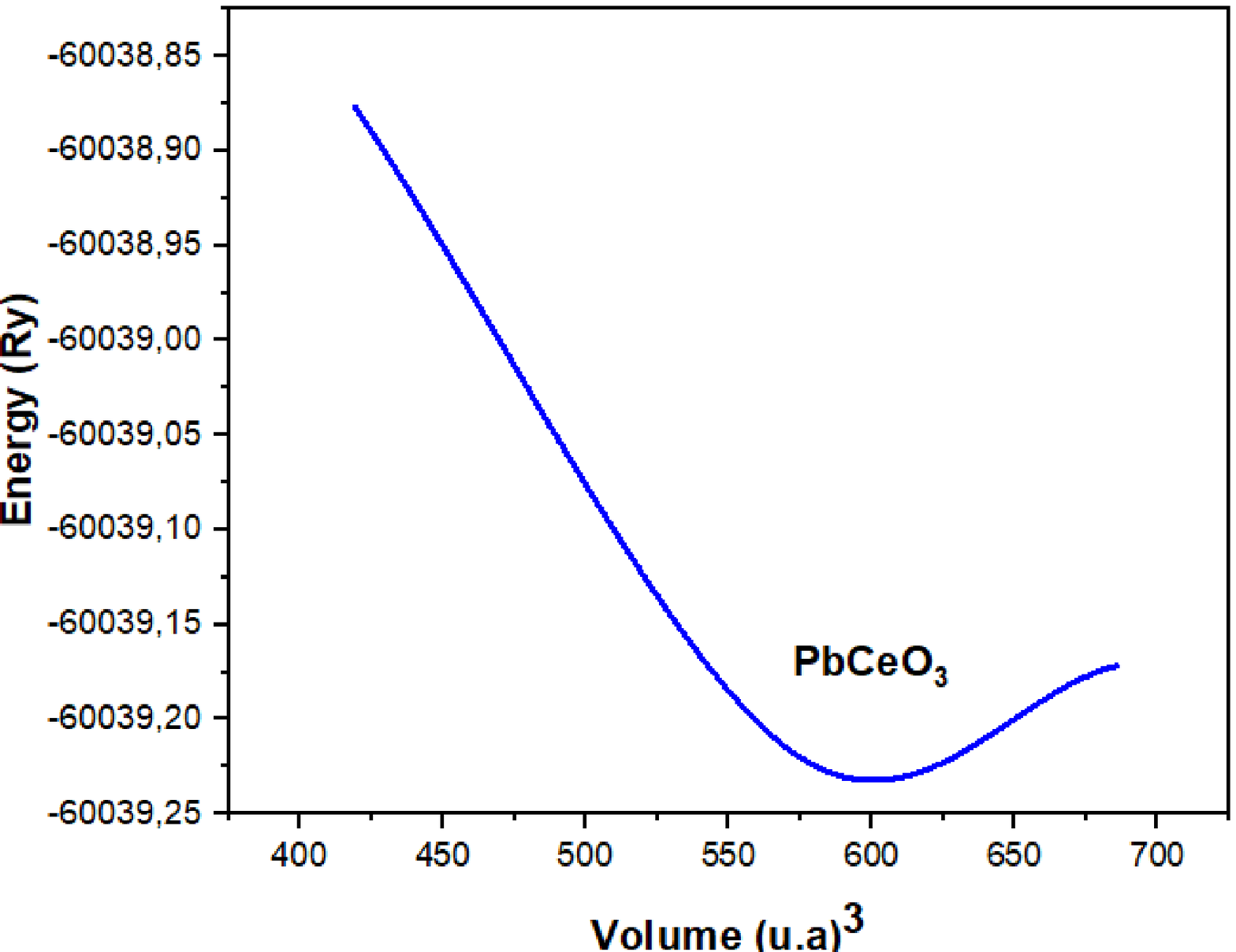}
	\includegraphics[scale=0.31]{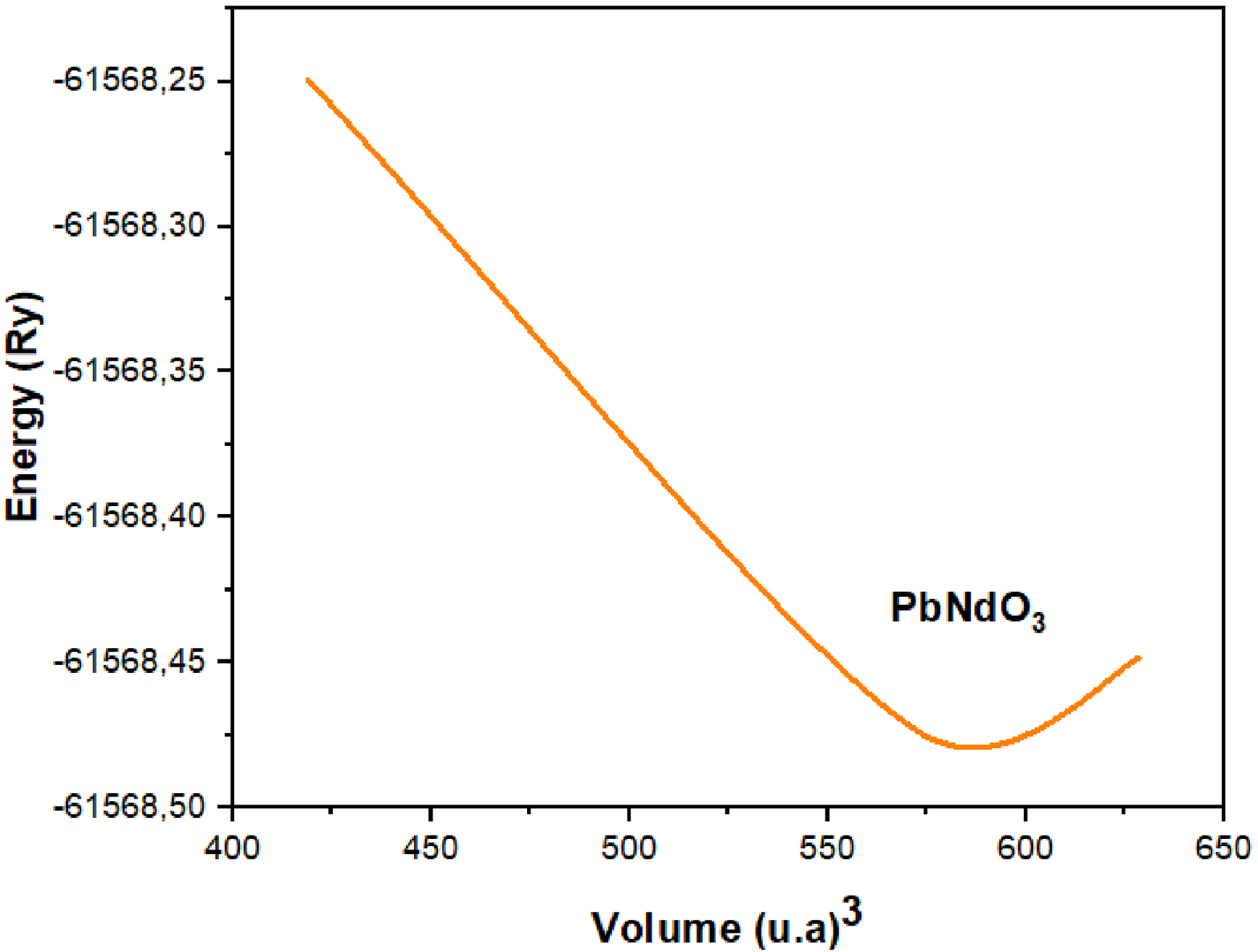}
	\includegraphics[scale=0.31]{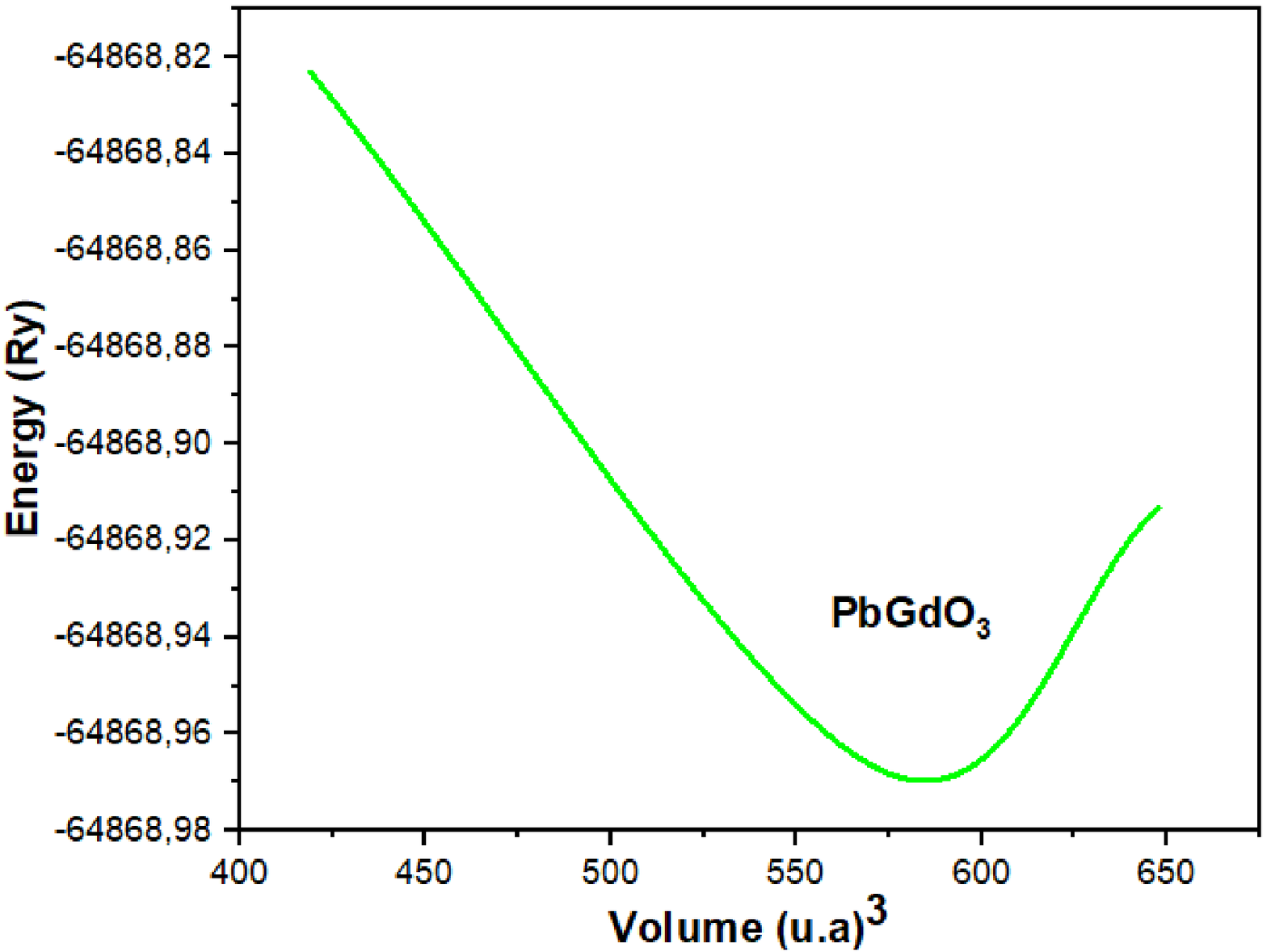}	
	\caption{The optimization plots of cubic $PbREO_3\; (RE=Ce,\; Nd\; and \;Gd)$ using PBE-GGA method.} \label{Figure:2}
\end{figure}

Perovskite oxides $PbREO_3\; (RE=Ce,\; Nd\; and \;Gd)$ have been studied in the cubic model (Pm-3m). The optimization of structural parameters through the unit cell volume was made to obtain the lattice constants, pressure derivative and bulk modulus. $E_0$ is taken as the minimum energy which is the ground state energy corresponding to the equilibrium volume. The extracted volume depending on the calculated energies are presented in figure \ref{Figure:1} by exploitation of Birch–Murnaghan’s low.
\begin{equation}
	E(V)= E_{0} +\dfrac{9V_{0}B}{16} \left\lbrace{\left\lbrace \left(\dfrac{V_{0}}{V}\right) ^{2/3} - 1\right\rbrace}^{2} B^{'} + {\left\lbrace \left( \dfrac{V_{0}}{V}\right) ^{2/3} - 1\right\rbrace}^{2}{\left\lbrace 6 - 4\left( \dfrac{V_{0}}{V}\right) ^{2/3}\right\rbrace}\right\rbrace.
\end{equation}
where $V_0$ presents the volume of the unit cell. $B$ is the bulk modulus with pressure derivative $B'$, related to the volume of the unit cell. Moreover, the variations of the total energy with respect of the unit cell volume are displayed in figure \ref{Figure:2}. We can see from the presented data an unequivocal relation between the a values and the size of anions. These values vary with the ionic radii of the considered rare-earth metals.
Table \ref{Table:2} summarizes the calculated values of lattice constant $a$, equilibrium volume $V_0$, the ground state energy $E_0$, bulk modulus $B$ and its derivative $B'$ for each material. \\
Experimental or theoretical investigation of structural properties have not been reported for our exciting materials. We have included the structural properties of cubic $BaCeO_3$ in order to compare them with our compounds \cite{15}, $BaCeO_3$ presents the same formula and behaves relatively as the same materials studied.
It is clear that the lattice parameters of the perovskites decrease slightly with increasing atomic numbers of the REs. The parameters $a$ decrease from Cerium to Gadolinium, and this is due mainly to the increases of atomic numbers. The extracted $a(\AA)$ for these materials are relatively compared with the experimental data's\cite{7,23}, and closely to previous calculations \cite{15}. The most stable cubic structure is shown for $Gd$ which can present a larger binding energies compared with $Nd$ and $Ce$. All studied compounds present a small pressure bulk modulus with a large volume, which indicate that they are more stable \cite{24,25}.
\begin{table}[H]
	\begin{center}
		\begin{tabularx}{1\textwidth} { 
				>{\raggedright\arraybackslash}X 
				>{\centering\arraybackslash}X
				>{\centering\arraybackslash}X
				>{\centering\arraybackslash}X
				>{\centering\arraybackslash}X
				>{\centering\arraybackslash}X
				>{\raggedleft\arraybackslash}X  }
			\hline 
			Materials  &Methods        & $a$(\AA) & $V_0$(\AA$^{3})$  & $B(GPa)$  & $B'$   & $E_{0}(Ry)$ \\
			\hline 
			PbCeO$_3$  &GGA   \vspace{0.2cm}& 4.4271    & 86.7677         & 104.0466  & 5.0000 & -60039.22 \\ 		
			PbNdO$_3$  &GGA    \vspace{0.2cm}& 4.3579    & 82.7622         & 106.4727  & 5.0000 & -61568.47 \\
			PbGdO$_3$  &GGA    \vspace{0.2cm}& 4.2930    & 79.1193         & 109.2162  & 5.0000 & -64868.98 \\
			
			BaCeO$_3$ &GGA\cite{15}     \vspace{0.01cm}& 4.4729    & 89.4886         & 110.2421  & 4.2689 & -34461.72 \\
			          &Exp\cite{23}    \vspace{0.1cm}& 4.4447    & 87.8066        &   &   & \\ 		
			\hline
		\end{tabularx}
		\caption {Calculated lattice constant $a$, volume $V_0$, Bulk modulus $B$, its primitive $B'$ and the total ground state energy $E_0$ of $PbREO_3\; (RE=Ce,\; Gd,\; Nd)$ compounds.}
		\label{Table:2}
	\end{center}
\end{table}
\subsection{Elastic properties}
The elastic properties are an important feature of solids which play a key role in materials science and novel technologies \cite{19}. They are used to describe the mechanical behaviors of materials in different situations. The elastic properties of perovskite oxides $PbREO_3\; (RE=Ce,\; Nd,\; Gd)$ are obtained by utilizing the ElaStic1.1 package \cite{19}.\\
For cubic crystalline structures, there are three independent elastic parameters which are denoted by $C_{11}$, $C_{12}$ and $C_{44}$. The mechanical stability is checked using Born criteria that are given by \cite{26,27}:
\begin{equation}
	C_{11} + 2C_{12} >0\;, \;\;C_{11} - C_{12} >0\;\; and \;\;	C_{44} >0.
\end{equation}
The elastic constants of cubic $PbREO_3\; (RE=Ce,\; Nd,\; Gd)$ are presented in table \ref{Table:3} and they respect Born's criteria which ensure that all these compounds are stable mechanically in the cubic structure. It is shown that the calculated value of $C_{11}$ is higher than $C_{12}$. This result indicates that the incompressibility along the crystallographic a-axis is stronger than along the b-axis. This means that the bonding strength along the [100] directions is higher compared with [011] direction in these compounds \cite{28}.
\begin{table}[H]
	\begin{center}
		\begin{tabularx}{1\textwidth} { 
				>{\raggedright\arraybackslash}X 
				>{\centering\arraybackslash}X
				>{\centering\arraybackslash}X
				>{\centering\arraybackslash}X
				>{\raggedleft\arraybackslash}X}
			\hline	
			Parameters              &$PbCeO_3$   & $PbNdO_3$& $PbGdO_3$  \\
			\hline     
			$C_{11}$ \vspace{0.2cm} & 197.9      & 208.1    & 218.2    \\
			$C_{12}$ \vspace{0.2cm} & 70.3       & 86.2     & 89.1     \\
			$C_{44}$ \vspace{0.2cm} & 31.8       & 41.5     & 48.3     \\	
			\hline
		\end{tabularx}
		\caption {The calculated elastic constants (in GPa) of $PbREO_3\; (RE=Ce,\;Gd,\;Nd)$ compounds.}
		\label{Table:3}
	\end{center}
\end{table}
From the calculated elastic constants $C_{ij}$, other mechanical constants such as shear modulus $G$, bulk modulus $B$, Cauchy pressure $C^{''}$, Pugh's ratio $B/G$, Poisson ratio $\nu$ and anisotropy factor $A$ are found through the Voigt-Reuss-Hill approximation \cite{29} using the following formulas \cite{19,30,31}:
\begin{equation}
	B=\dfrac{C_{11}+2C_{12}}{3} \hspace*{0.5cm} G=\dfrac{G_{V}+G_{R}}{2} \hspace*{0.5cm}
	Y=\dfrac{9GB}{G+3B}\hspace*{0.5cm} \nu=\dfrac{3B-2G}{2(3B+G)}.
\end{equation}
where $G_{V}$ and $G_{R}$ are the shear modulus of Voight and Reuss approaches, respectively.
\begin{equation}
	G_{V}=\dfrac{C_{11}-C_{12}+3C_{44}}{5}\; \;
	G_{R}=\dfrac{5C_{44}(C_{11}-C_{12})}{4C_{44}+3(C_{11}+C_{12})}
\end{equation}
The calculated results listed in table \ref{Table:4} show that the bulk modulus for $PbGdO_3$ is more larger than the others which signifies that the changes in volume withstand more from all changes when compressed from all sides. The hardness or stillness behavior of any material is determined by the mechanical parameter Young's modulus $Y$ \cite{32}. 
We noted that the higher Young's modulus value of $143.09\; GPa$ is observed for $PbGdO_3$ which means that $PbGdO_3$ is a rigid structure compared to others, and it confirms the previous results found. 
The Pugh's ratio $B/G$ shows the brittle or ductile aspect of materials. The critical value of the $B/G$ ratio which distinguishes between ductile and brittle is $1.75$. A low $B/G$ value of $1.75$ is associated with brittleness aspect, while a high one has consisted of ductility \cite{32}. The obtained values clearly indicate that the $PbREO_3\; (RE=Ce,\; Nd,\; Gd)$ compounds have the Pugh's  ratio highest than the critical value of $1.75$ which indicates that all these perovskites behave as ductile materials. In addition, Cauchy pressure can be used also to show the brittle and ductile nature of a material. It is defined as the difference between the elastic constant $C_{12}-C_{44}$ \cite{28}. When the Cauchy pressure is negative, the material will be brittle, otherwise, it will be ductile. As shown in table \ref{Table:4}, the Cauchy pressure is positive for all perovskites, which confirms the ductile nature. The calculation of Poisson's ratio $\nu$ separates the brittle and ductile nature of solids \cite{28}. According to its role, the Poisson's ratio $\nu$ values confirm again the previous results.
\begin{table}[H]
	\begin{center}
		\begin{tabularx}{1\textwidth} { 
				>{\raggedright\arraybackslash}X 
				>{\centering\arraybackslash}X
				>{\centering\arraybackslash}X
				>{\centering\arraybackslash}X
				>{\raggedleft\arraybackslash}X  }
			\hline
			Parameters                     & $PbCeO_3$  &$PbGdO_3$ & $PbNdO_3$  \\
			\hline     
			$B$         \vspace{0.2cm}     & 112.87     & 132.12   & 126.84   \\
			$G$         \vspace{0.2cm}     & 42.19      & 54.22    & 48.40   \\
			$Y$         \vspace{0.2cm}     & 112.54     & 143.09   & 128.81   \\
			$B/G$       \vspace{0.2cm}     & 2.68       & 2.43     & 2.62   \\
			$C^{''}$    \vspace{0.2cm}     & 38.50      & 40.80    & 44.70   \\
			$\nu$       \vspace{0.2cm}     & 0.33       & 0.33     & 0.32   \\		
			$A$         \vspace{0.2cm}     & 5.711      & 1.012    & 1.764   \\
			\hline
		\end{tabularx}
		\caption {Calculated values of bulk modulus $B$ , shear modulus $G$, Young’s modulus $Y$, Pugh’s ratio $B/G$, Cauchy pressure $C^{''}$, Poisson’s ratio $\nu$ and elastic anisotropic factor $A$ of $PbREO_3\; (RE=Ce,\;Nd,\;Gd)$ compounds.}
		\label{Table:4}
	\end{center}
\end{table}
Finally, we proceed to calculate Anisotropy of all these systems which is an important parameter used to describe the directional variation depending on properties \cite{28}. The anisotropic factor's calculated values are 5.711, 1.042 and 1.765 for $PbCeO_3$, $PbGdO_3$ and $PbNdO_3$, respectively. $PbCeO_3$ is then shown as more anisotropic than the others, and the $PbGdO_3$ presents a nearly equal unity indicating a close to isotropic nature.
\subsection{Electronic properties}
\begin{figure}[H]
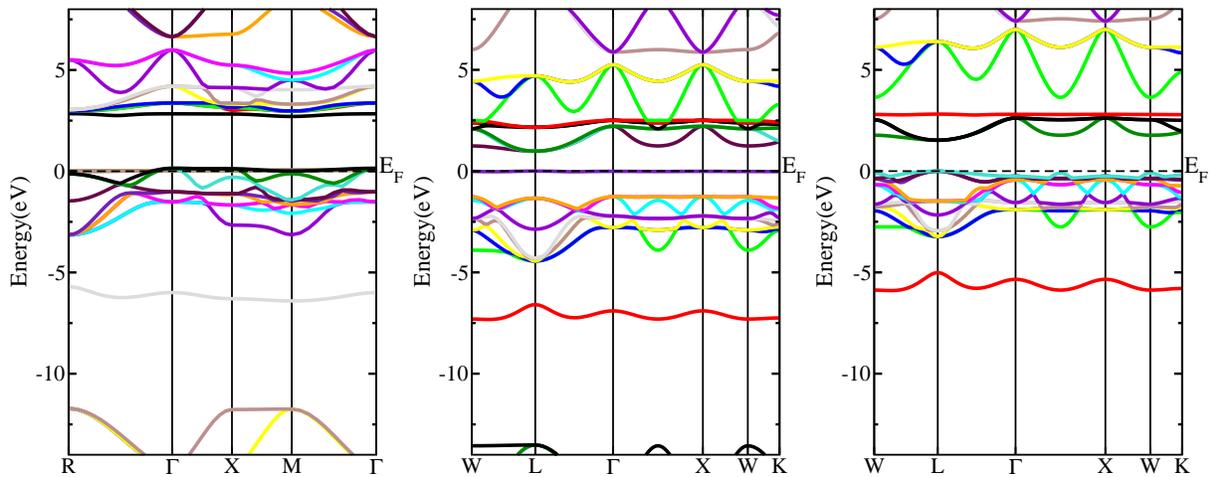

	\centering
	\includegraphics[scale=0.25]{band_Ce.eps} 
	\includegraphics[scale=0.25]{band_Nd.eps}
	\includegraphics[scale=0.25]{band_Gd.eps} 	
	\caption{Band structure of $PbREO_3\; (RE=Ce,\;Nd,\;Gd)$ compounds.} 
	\label{Figure:3}
\end{figure}

To investigate the electronic properties of cubic $PbREO_3\; (RE=Ce,\;Nd,\;Gd)$ perovskite oxides, we have calculated the band structure, the total and partial density of states. We have used GGA-PBE correlation energy functional in the DFT method. 
The fundamental band gap energies are found to be $1.74\;eV$, $0\;eV$ and $0\;eV$ for $PbCeO_3$, $PbNdO_3$ and $PbGdO_3$, respectively.\\
In order to remedy the problem of underestimation of the gap energy, we have utilized the new-mBJ approach which will improve very precisely the electronic band gaps, especially in semiconductors and insulators \cite{20}. Along the high directions in the first Brillouin zone, we have calculated the electronic band structures of $PbREO_3\; (RE=Ce,\;Nd,\;Gd)$ as illustrated in figure \ref{Figure:3}.  We notice that the gap energy between the conduction band minimum (CBM) and the valence band maximum (VBM) of $PbCeO_3$ are placed at $\Gamma$ point which means that $PbCeO_3$ has a direct band gap equal to $2.60\;eV$.  However, the CBM and VBM for both $PbGdO_3$ and $PbNdO_3$ perovskites are sited at the same symmetry point $L$, and this suggests that $PbGdO_3$ and $PbNdO_3$ have a direct band gaps which are equal to $1.55\;eV$ and $1.02\;eV$, respectively. These obtained results by using new-mBJ approach are in full agreement with previous theoretical and experimental works, where the gap energies are closed and compared with RE-based perovskite oxides \cite{6, 15}. Around Fermi level energy, the movement of bands occupied by electrons are the main responsible of the appeared gap energy for all systems studied. The direct transitions shown in the figure above, will be detailed by a calculation of PDOS.
To analyse the contribution of different band energies in the electronic states, we have studied the total (TDOS) and partial density of states (PDOS) of $PbREO_3\; (RE=Ce,\; Nd,\; Gd)$ which are presented in figure \ref{Figure:4}. The Fermi level $E_f$ is set at $0\;eV$. \\ 
As for each perovskite system, different energy bands containing several energy levels appear in the studied structures. In the present analysis, we will study the effect that rare earth elements can bring on the cubic structure $PbREO_3$. Starting with $RE=Ce$, the proven opening gap in the band structure is essentially due to the contribution of the $2p$ orbitals of oxygen and $f$ of $Ce$. The conduction band is ensured by the presence of an intense band due to the $f$ of $Ce$ and $p$ of $Pb$, and the valence band is essentially due to the $p$ of oxygen. 
In the conduction band, Neodymium contributes with $f$ occupied orbital states, and they show a weak hybridization with $p$ states of oxygen. The $p$ orbitals of $Pb$ present occupied states ensuring an important band in the total density. The $p$ of oxygen contributes with a strong intra-band within the valence band and this from $-5\;eV$ to $0\;eV$. For $PbGdO_3$, the $f$ occupied orbital states of $Gd$  show a weak hybridization with $p$ states of oxygen in the CB, and they also present an important band in the total density. However, the $p$ orbitals of oxygen have a major contribution in the valence band.\\
The overall spectra of TDOS and PDOS are almost similar in all our systems. The valence band (VB) can be arranged into two regions. The lower region of the band placed before $-5\;eV$ consists of the ione pairs of $Pb-6s^2$ and presents a large dispersion which suggests a covalent bonding with Oxygen. This result is also shown for the case of $PbMnO_3$ \cite{33}. The upper region between $-5$ to $0\;eV$ (Fermi level) is mainly contributed by $O-2p$ and $4f-REs$ orbitals, which shows a weak hybridization for $PbCeO_3$ and $PbNdO_3$ and a strong hybridisation of $4f-Gd$ and $2p-O$ for $PbGdO_3$. The $4f-REs$ orbitals show the gap energy of all these compounds with an observed hybridization due to $p$ and $f$ orbitals of Oxygen and $REs$ respectively \cite{15}. 
For the CB, a hybrid state between the $p$ orbitals of $Pb$ and Oxygen could be observed in the $4-5.2\; eV$, $3-5\; eV$ and $4-7.5\; eV$ energy ranges for $PbCeO_3$, $PbNdO_3$ and $PbGdO_3$, respectively.
\begin{figure}[H]
	\includegraphics[scale=0.55]{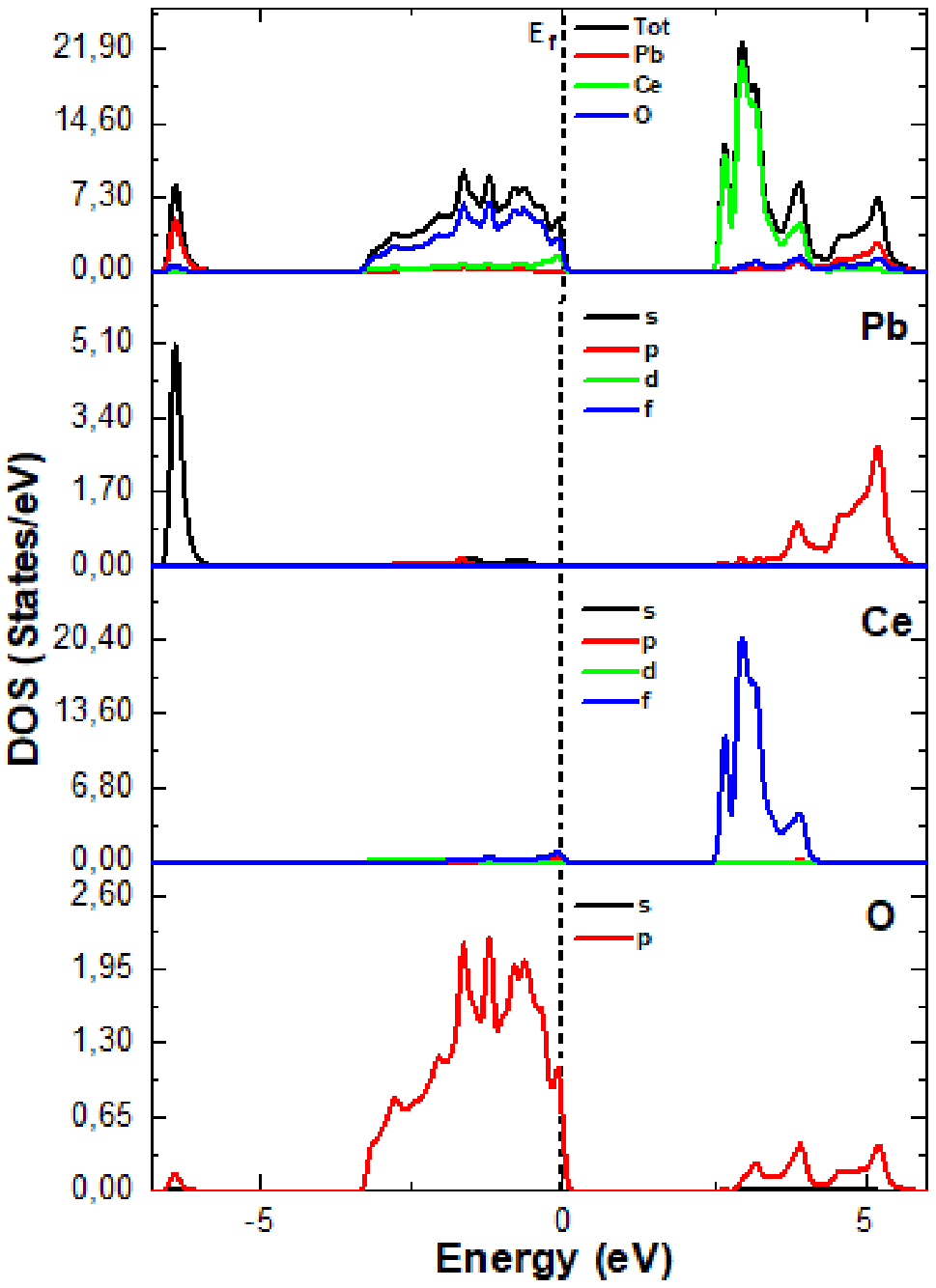} 
	\includegraphics[scale=0.55]{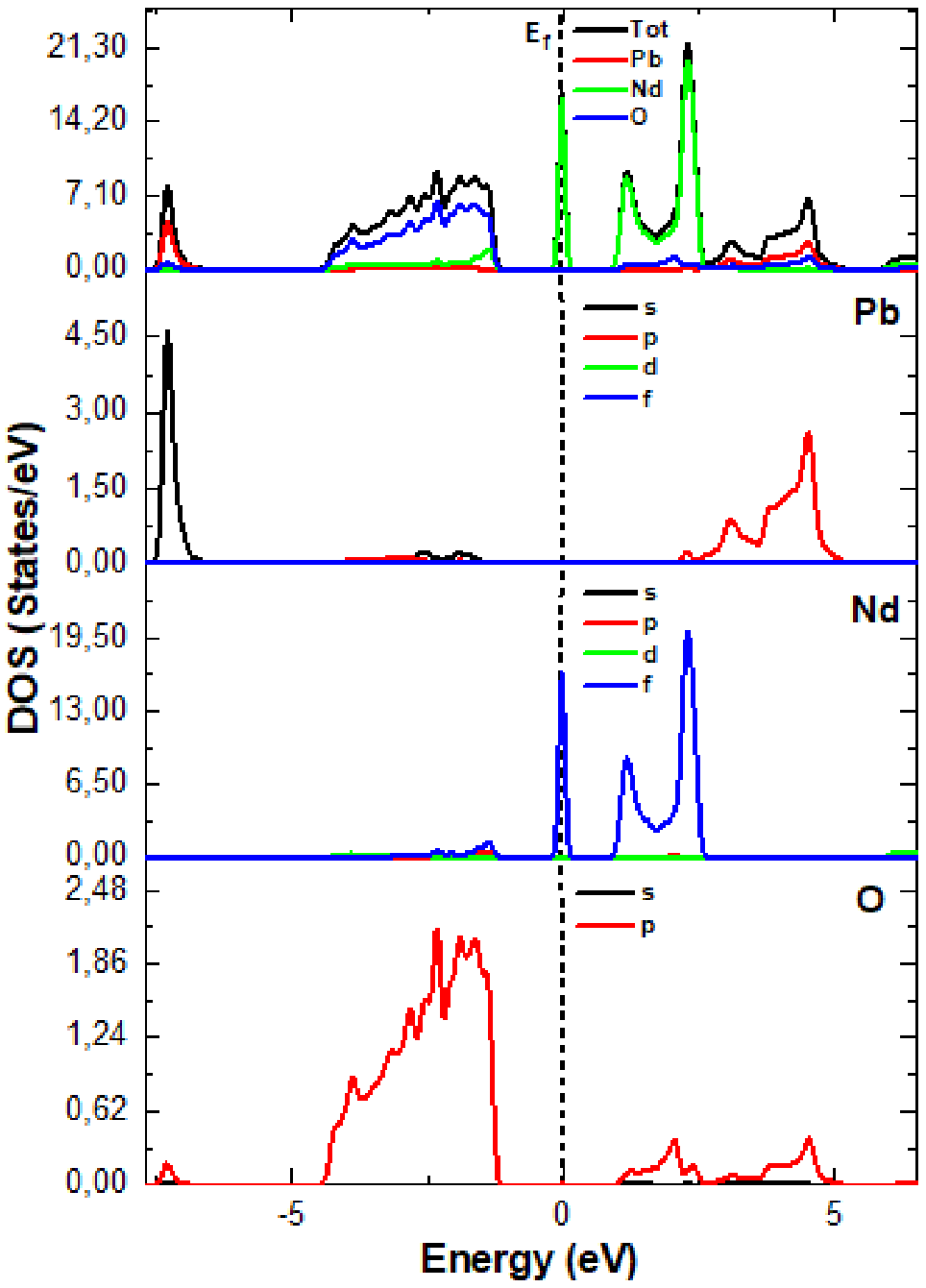} 	\includegraphics[scale=0.55]{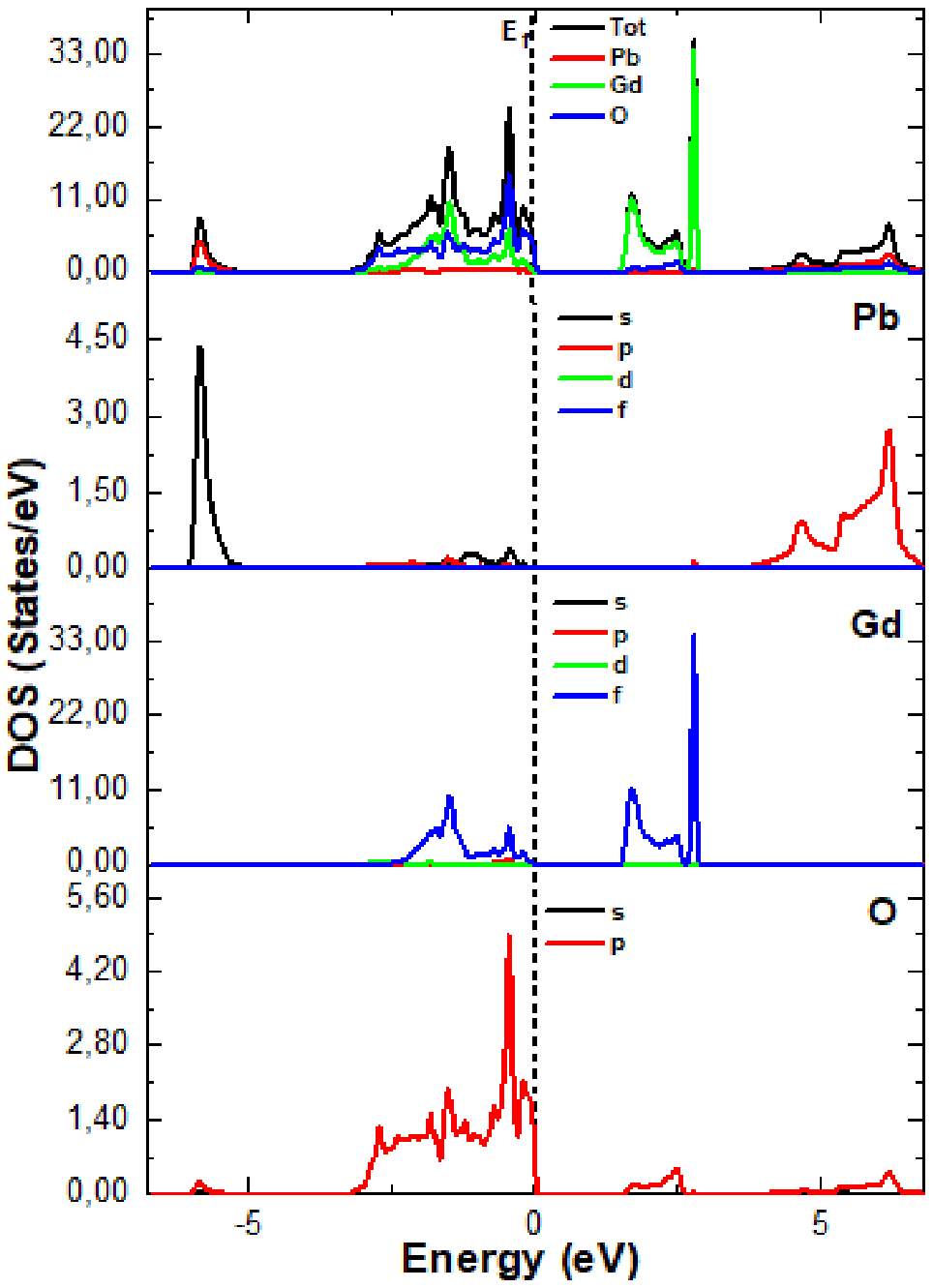}
	\caption{The partial and total density of the states of $PbREO_3\; (RE=Ce,\;Gd,\;Nd)$ perovskite oxides.} \label{Figure:4}
\end{figure}
\subsection{Optical properties}
In view of understanding and exploring the optical characteristic of $PbReO_3\; (Re=Ce,\;Nd,\;Gd)$, we have studied the optical properties such as the dielectric function, absorption coefficient, refraction index and reflectivity. All these optical properties are found from the dielectric function which is identified by $\varepsilon (\omega)=\varepsilon_{1} (\omega) + i\varepsilon_{2} (\omega)$, through its real $\varepsilon_{1} (\omega)$ and imaginary $\varepsilon_{2} (\omega)$ parts \cite{34,35}.\\
The imaginary part $\varepsilon_{2} (\omega)$ is given by the following equation \cite{34,35}:
	\begin{equation}
	\varepsilon_{2} (\omega) = \dfrac{4\pi^{2}e^{2}}{m^{2}\omega^{2}} \sum\limits_{n,m}\int_{k} |M_{nm}|^{2} f_{n}(1-f_{n}) \delta (E_{mk}-E_{nk}-w) d^{3}k,
\end{equation}
where $M_{nm}$ presents the transition matrix element. $e$ and $m$ denote successively the charge and mass of free electron. The real part can be expressed as follows \cite{34, 35}:
\begin{equation}
	\varepsilon_{1} (\omega) = 1 + \dfrac{2}{\pi}P \int_{0}^{\infty} \dfrac{\omega' \varepsilon_{2} (\omega') }{\omega'^{2} - \omega^{2} } d\omega
\end{equation} 
where $P$ is the principal value of the integral.\\

\begin{figure}[H]
	\centering
	\includegraphics[scale=0.7]{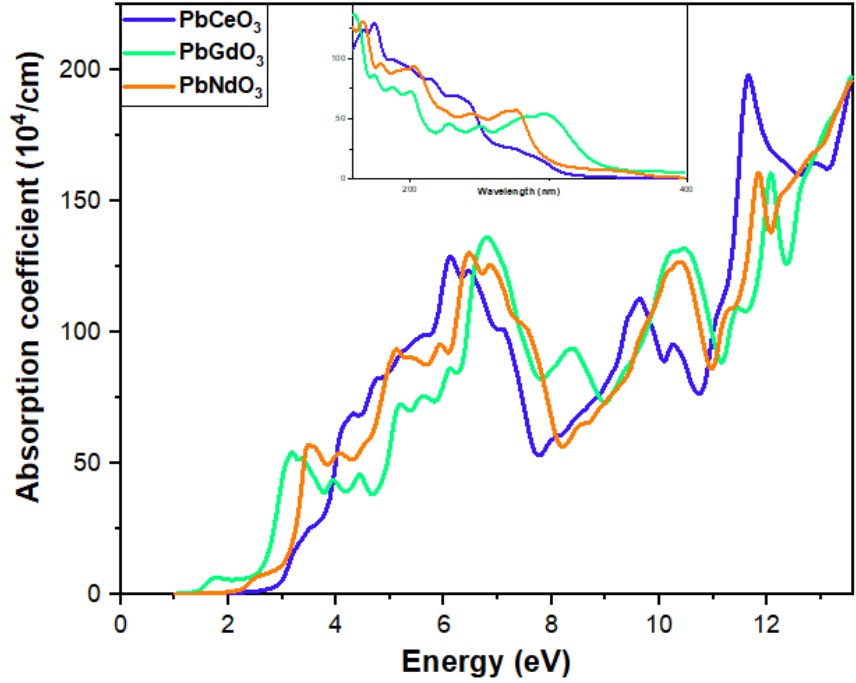}
	\caption{Calculated absorption coefficients spectra of cubic $PbReO_3\; (Re=Ce,\;Nd,\;Gd)$ perovskite oxides using nmBJ-GGA potential. } \label{Figure:5}
\end{figure}

\begin{figure}[H]
	\centering
	\includegraphics[scale=0.5]{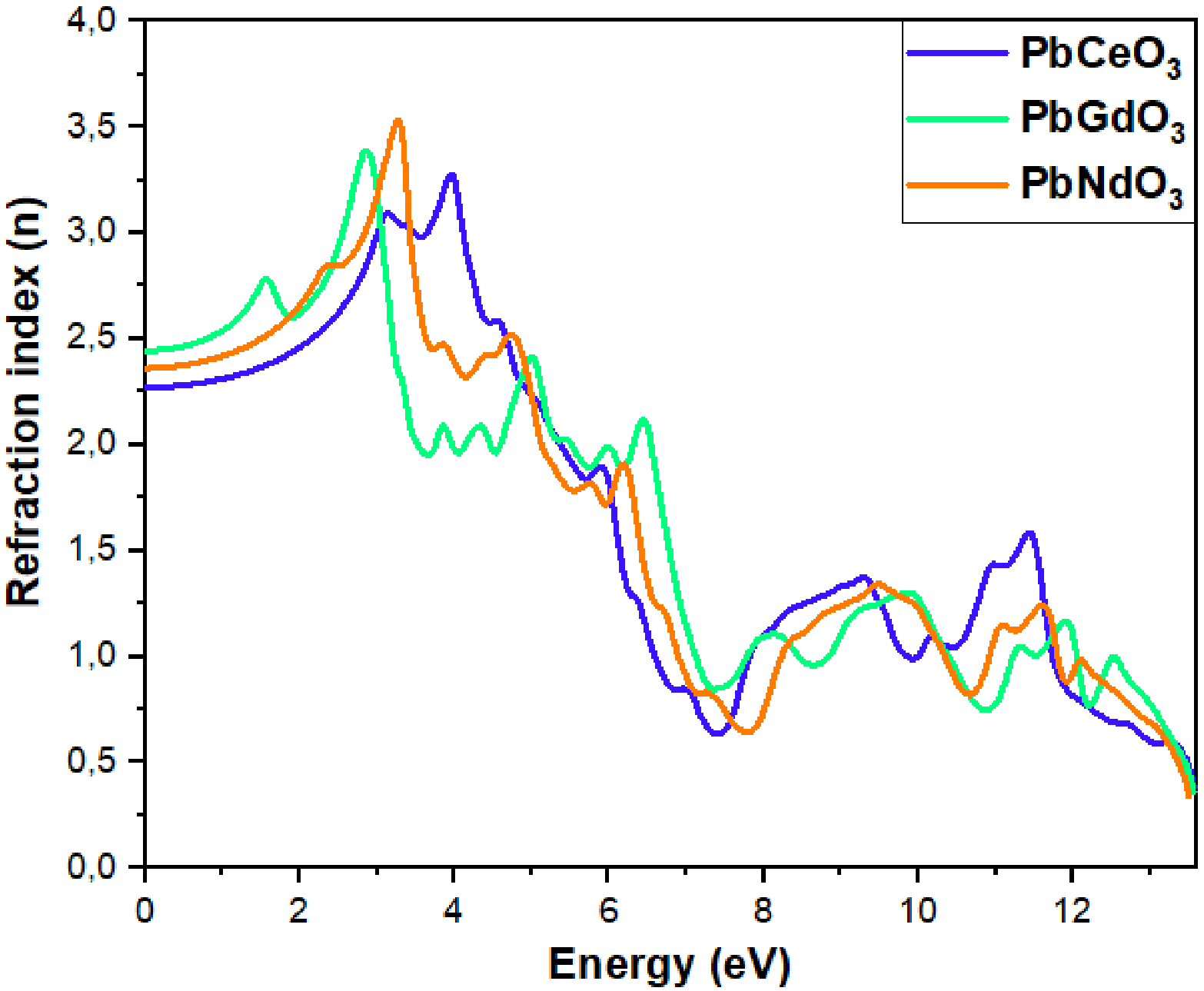}
	\includegraphics[scale=0.51]{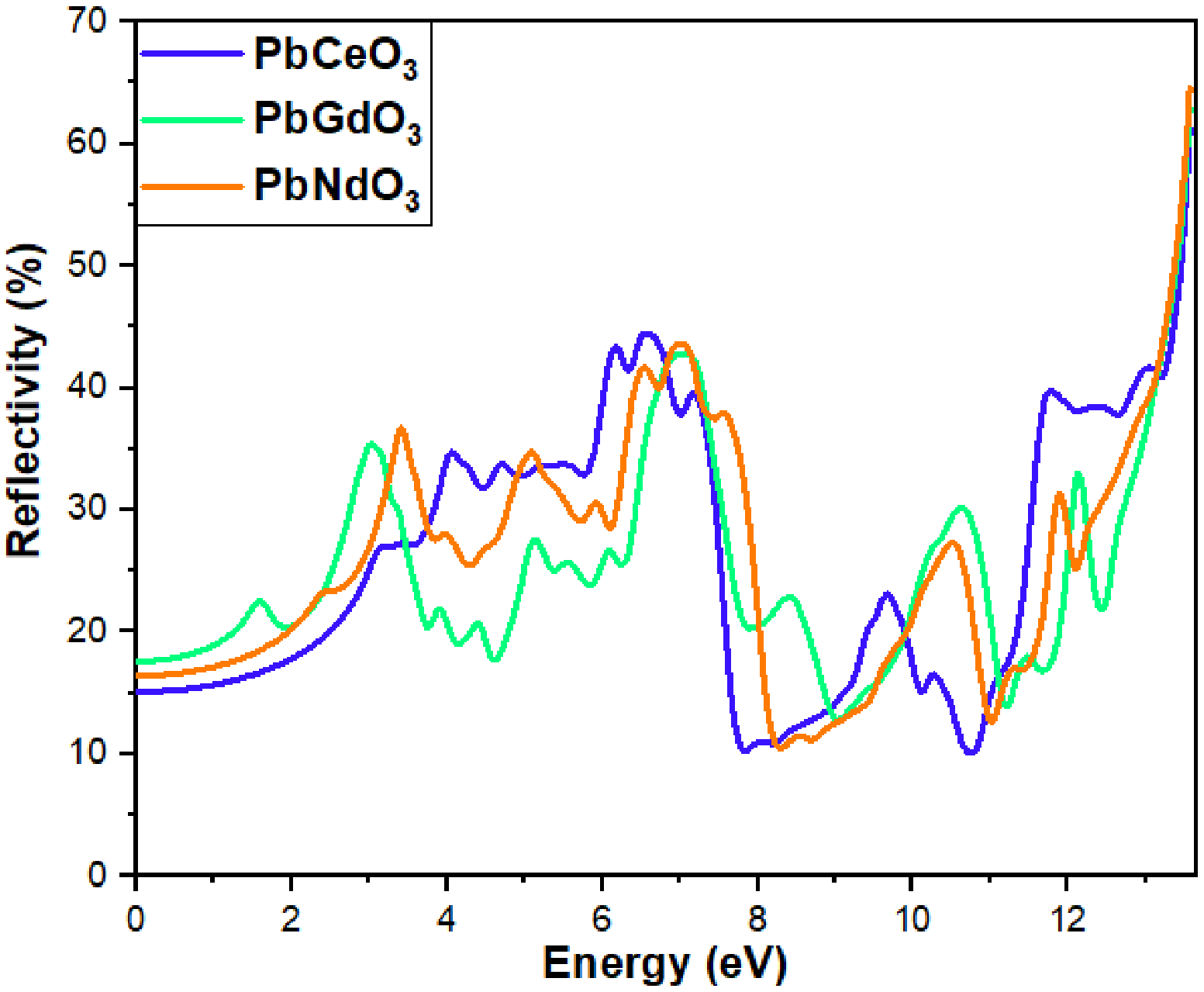}
	\caption{Calculated the refractive index $n(\omega)$ and reflectivity $R(\omega)$ spectra of cubic $PbReO_3\; (Re=Ce,\;Nd,\;Gd)$ perovskite oxides using nmBJ-GGA potential.} \label{Figure:6}
\end{figure}

The absorption coefficient $\alpha(\omega)$, the reflectivity $R(\omega)$ and the refractive index $n(\omega)$ can be obtained using the following equations \cite{26,27}:
\begin{equation}
	\alpha(\omega)=\dfrac{\omega \epsilon_{2}(\omega)}{c}
\end{equation}
\begin{equation}
	R(\omega)=\displaystyle\left\lvert \dfrac{\sqrt{\epsilon(\omega)-1}}{\sqrt{\epsilon(\omega)+1}}\right\rvert ^{2}
\end{equation}
\begin{equation}
	n(\omega)=\dfrac{1}{\sqrt{2}} \sqrt{\sqrt{\epsilon_{1}^{2}(\omega)+\epsilon_{2}^{2}(\omega)}+\epsilon_{1}(\omega)}
\end{equation}
The obtained results for the imaginary and real parts of the dielectric function $\epsilon(\omega)$ corresponding on photon energy are shown in figure \ref{Figure:4}(a,b). It is well known that the real part of the dielectric function $\epsilon_{1}(\omega)$ is described as the dispersion of incident photon by the material \cite{36}. Whereas, the imaginary part corresponds to optical transitions from the conduction band to the valence band and it is linked to the energy absorbed by the material \cite{36}.\\
The static dielectric constant $\epsilon_{1}(0)$ is determined by the low energy of $\epsilon_{1}(\omega)$. The calculated values of $\epsilon_{1}(0)$ of all these compounds are presented in table \ref{Table:5}. It is seen in figure \ref{Figure:4} (a) that $\epsilon_{1}(\omega)$ increases from $\epsilon_{1}(0)$ and reaches a maximum values of $3.90\;eV$, $2.81\;eV$ and $3.27\;eV$ respectively, for $PbCeO_3$, $PbGdO_3$ and $PbNdO_3$, and then it decreases continuously and became negative. Such a negative $\epsilon_{1}(\omega)$ indicates that all these compounds will behave as metals in this range and have a reflection of incident light on them. The imaginary part $\epsilon_{2}(\omega)$ of $PbReO_3\; (Re=Ce,\;Nd,\;Gd)$ is depicted in figure \ref{Figure:4} (b). It can be remarked that $\epsilon_{2}(\omega)$ values are completely null at low photons energies. The critical points of $\epsilon_{2}(\omega)$ are found approximately at $2.60\;eV$, $1.55\;eV$ and $1.02\;eV$ for $PbCeO_3$, $PbGdO_3$ and $PbNdO_3$ respectively, which were nearly related to optical band gaps of each material. 
In addition to that, the maximum values are attained in the UV-region, this explained by a great and important absorption. On the other hand, in the particular visible range, we notice that $PbGdO_3$ presents high and large absorption compared to $PbCeO_3$ and $PbNdO_3$.
The calculated imaginary and real parts of the dielectric function of our compounds are in good agreement with the obtained results in other studies \cite{15,37}.
\begin{figure}[H]
	\centering
	\includegraphics[scale=0.5]{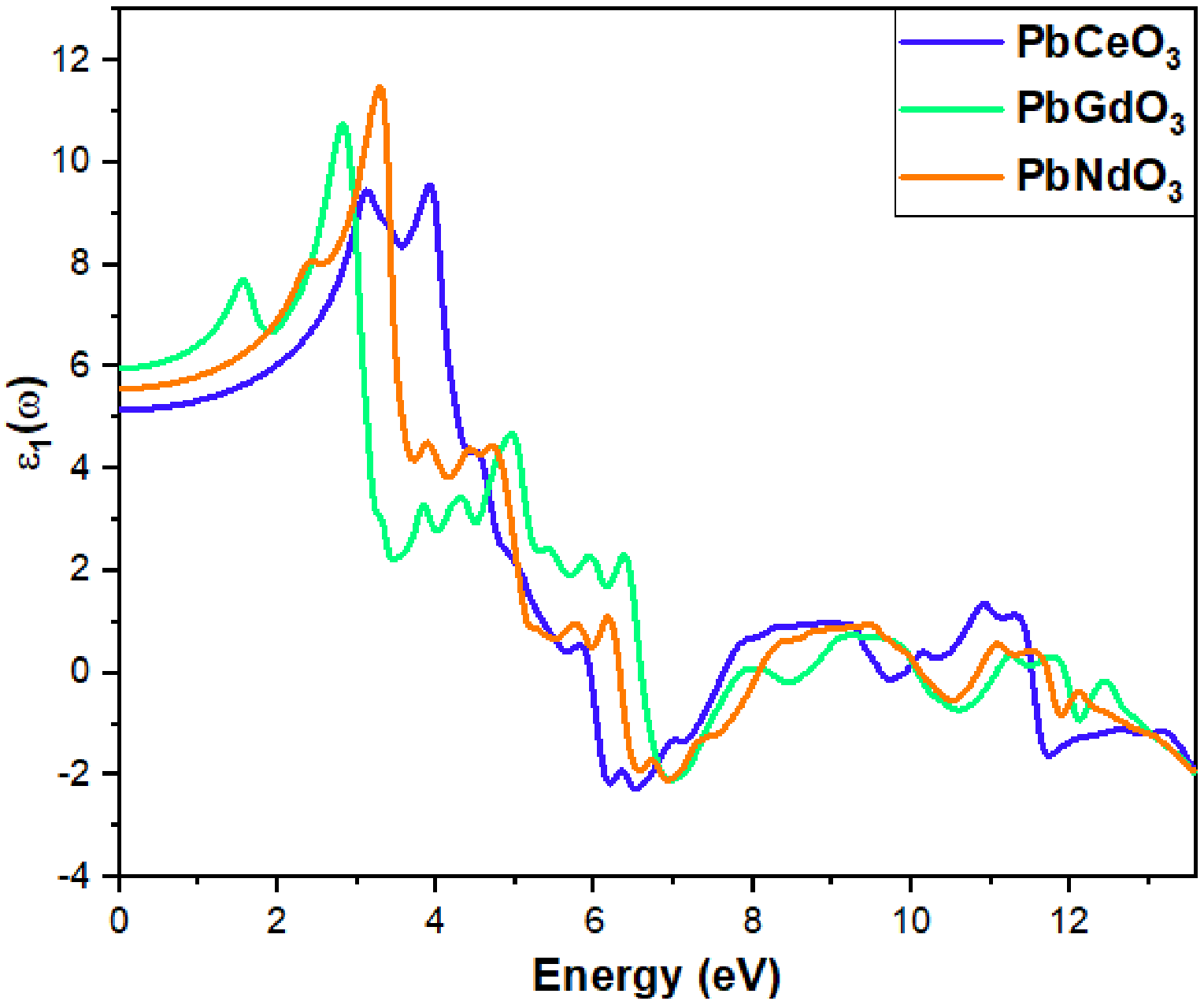}
	\includegraphics[scale=0.5]{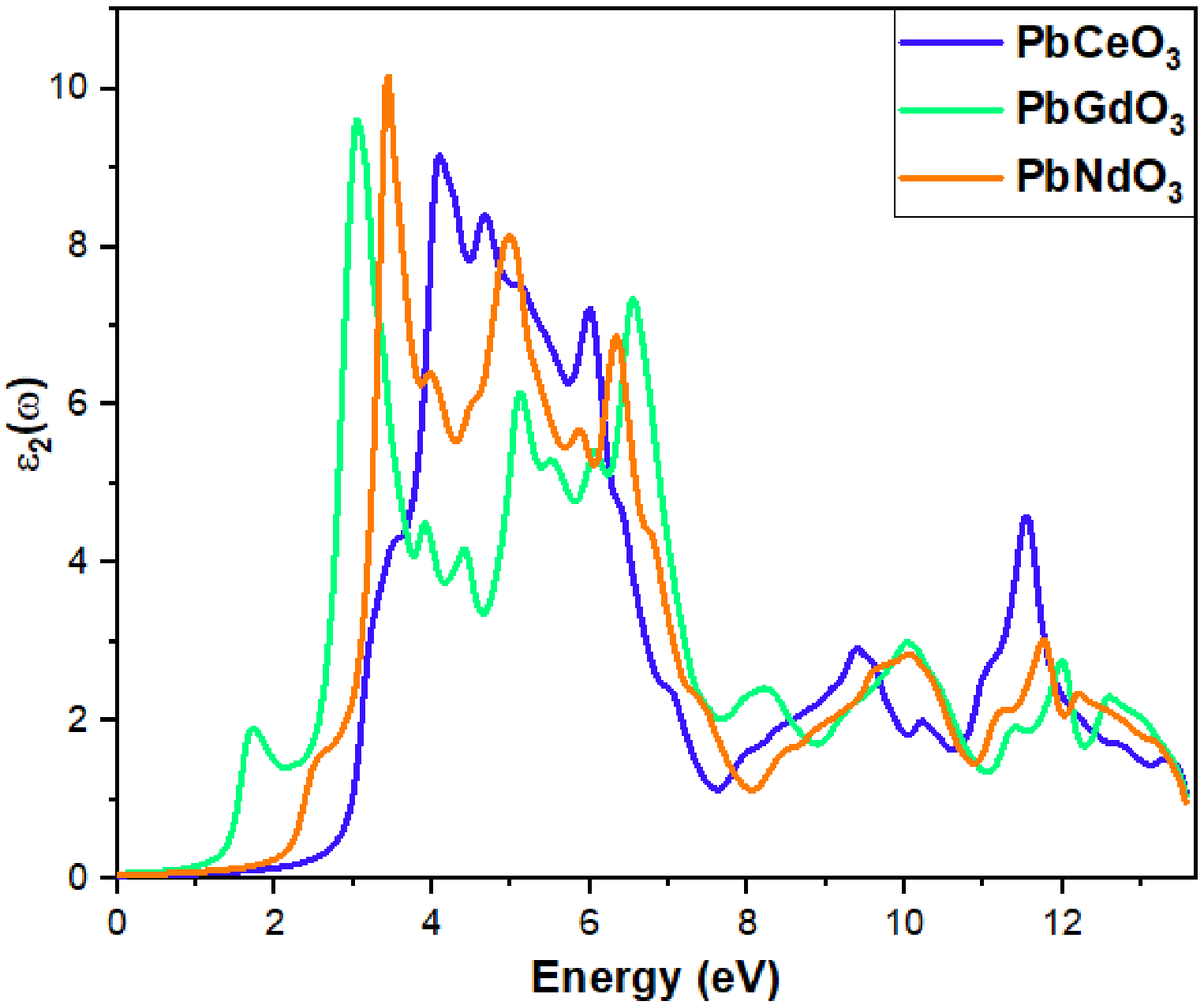}
	\caption{Calculated real part and imaginary part dielectric constant of cubic $PbReO_3\; (Re=Ce,\;Nd,\;Gd)$ perovskite oxides using nmBJ-GGA potential. } \label{Figure:4}
\end{figure}
The absorption coefficient corresponds to the ability of a material to absorb the incident radiations \cite{36}. Thus, the absorption coefficient of all the compounds as a function of the energy varying from $0\;eV$ to $13\;eV$ and also in the function of the wavelength are illustrated in figure \ref{Figure:5}. We note that $\alpha(\omega)$ is closely related to $\epsilon_{2}(\omega)$ of the dielectric function. Critical points in the $\alpha(\omega)$ spectra are approximately $2.76$, $1.61$ and $1.80\;eV$ for $PbCeO_3$, $PbGdO_3$ and $PbNdO_3$, respectively. 
The absorption coefficients $\alpha(\omega)$ increases when the photon energy increases. Based on this aspect, these materials can be exploited as a promote candidates for various optoelectronic applications,
Figure \ref{Figure:6} (a, b) shows respectively the reflectivity $R(\omega)$ and the refraction index $n(\omega)$ as a function of energy. It is clear that the spectrum of $n(\omega)$ is closely like the real part of the dielectric function $\epsilon_{1}(\omega)$. 
This aspect is a result of the relation $n(0)=\sqrt{\epsilon_1(0)}$. The calculated values of $n(0)$ and $R(0)$ for $PbReO_3\; (Re=Ce,\;Nd,\;Gd)$ are presented in table \ref{Table:5}. 
Beyond the values of static refractive indices, $n(\omega)$ begin to increase, and its reaches the maximum values approximately of $3.28$, $3.40$ and $3.53$ for $PbCeO_3$, $PbGdO_3$ and $PbNdO_3$ at $3.95\;eV$, $2.87\;eV$ and $3.33\;eV$, respectively. These picks are explained by the low transmittance of all treated compounds. 
Then, the refractive index decreases at high energies, indicating that the high energy of photons are absorbed by the material. On the other hand, figure \ref{Figure:4} (b) presents reflectivity with respect to photon energy. 
With higher photon energies, we show that $R(\omega)$ behaves as  $\epsilon_{1}(\omega)$. We note that the reflectivity of each material has the maximum values of $2$ at $7\;eV$. In this energy zone, the materials can be employed as a screen because it reflects most of the electromagnetic waves, especially $PbCeO_3$.
These optical results of the studied compounds have similar behavior with several works \cite{15,37,38}.
\begin{table}[H]
	\begin{center}
		\begin{tabular}{|p{4.5cm}|p{3cm} p{3cm} p{1cm}|}
			\hline
			\diagbox{Compounds}{Parameters} & $\epsilon_1(0)$ & $n(0)$ & $R(0)$\\
			\hline \hspace{1.5cm} \vspace{0.1cm}
			$PbCeO_3$ & 5.16 & 2.27 & 0.15  \\
			\hline \hspace{1.5cm} \vspace{0.1cm}
			$PbGdO_3$ & 5.97 & 2.44 & 0.18  \\
			\hline \hspace{1.5cm} \vspace{0.1cm}
			$PbNdO_3$ & 5.57 & 2.36 & 0.16  \\
			\hline
		\end{tabular}
		\caption {Calculated of dielectric constant $\epsilon_1(0)$, refractive index $n(0)$ and reflectivity $R(0)$ of $PbReO_3\; (Re=Ce,\;Nd,\;Gd)$ perovskites.}
		\label{Table:5}
	\end{center}
\end{table}
\subsection{Thermoelectric properties}
Thermoelectric materials are one class of materials which can be employed to transform heat to electrical energy. Thermoelectric (TE) properties have been calculated close to the Fermi level, by exploiting the semi-classical Boltzmann theory as introduced in the BoltzTrap package \cite{21}. TE properties for cubic $PbREO_3\; (RE=Ce,\;Nd,\;Gd)$ compounds have been studied in relation to electronic relaxation time-dependent electrical conductivity ($\sigma / \tau$ ), Seebeck coefficient ($S$) and the power factor ($PF$) as a function of temperature. Electronic relaxation time ($\tau$) was used to calculate TE properties, which defines the time between two successive collisions of the charge carriers \cite{35,39}. \\
The calculated values of $\sigma / \tau$, $S$ and PF can be deduced based on relaxation time approximation from the following expressions \cite{40}:
\begin{equation}
	S_{\alpha\beta} (T, \mu) = \dfrac{1}{e^{2}T\Omega \sigma_{\alpha\beta} (T, \mu)} \int \sigma_{\alpha\beta} (\varepsilon) (\varepsilon-\mu) \left[\dfrac{\partial f_{\mu}(T, \mu)}{\partial \varepsilon} \right] d\varepsilon,
\end{equation}
\begin{equation}
	\sigma_{\alpha\beta} (T, \mu) = \dfrac{1}{\Omega} \int \sigma_{\alpha\beta} (\varepsilon) \left[\dfrac{\partial f_{\mu}(T, \mu)}{\partial \varepsilon} \right] d\varepsilon,
\end{equation}
\begin{equation}
	PF = \dfrac{S_{\alpha\beta}^{2}(T, \mu) \sigma_{\alpha\beta} (T, \mu)}{\tau},
\end{equation}
where $\Omega$ is the volume of the supercell, $f_\mu$ is the Fermi distribution and ($\alpha$, $\beta$) correspond the tensor indicators.
\subsubsection*{Seebeck coefficients:}
\begin{figure}[H]
	\centering
	\includegraphics[scale=0.95]{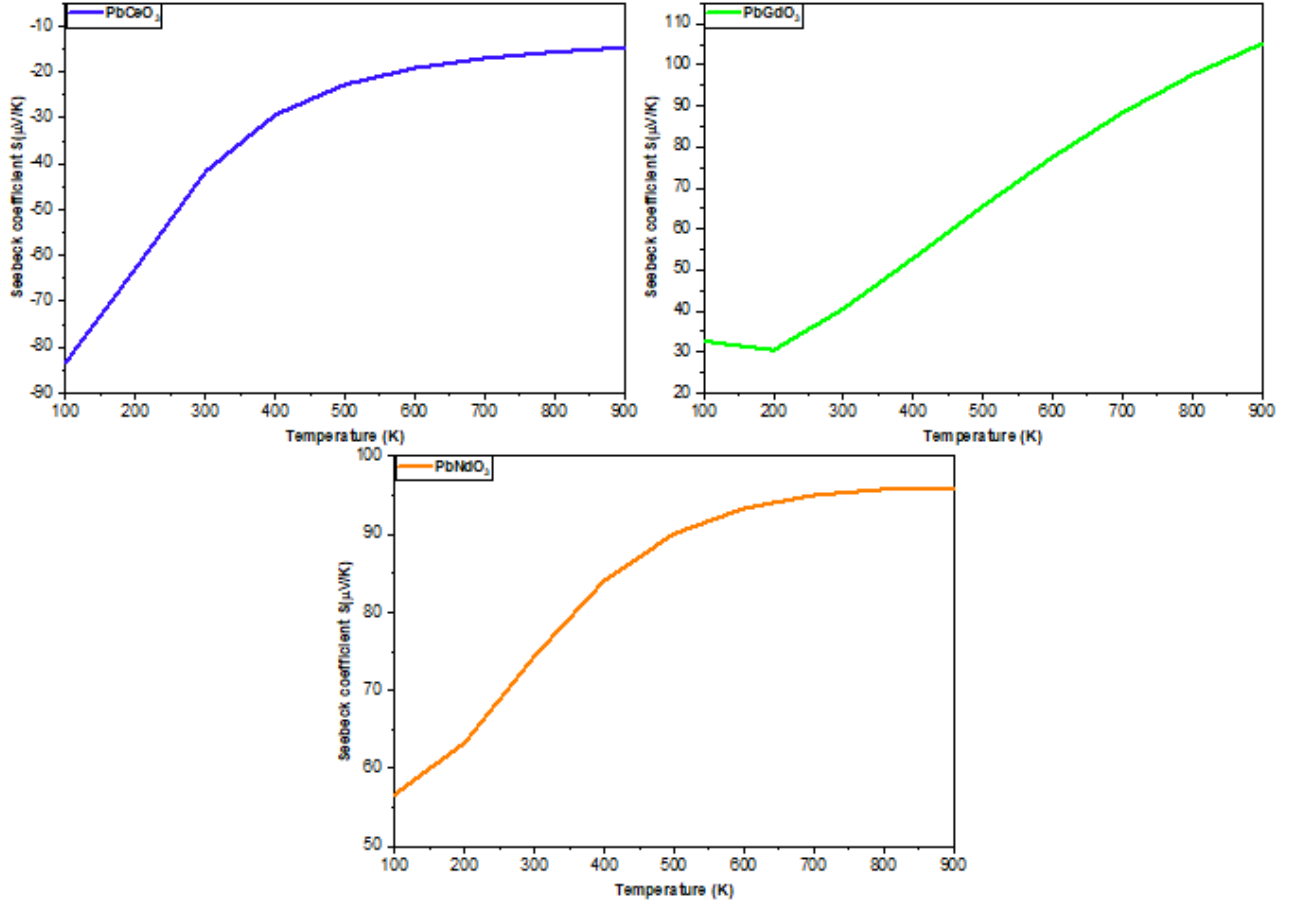}
	\caption{Seebeck coefficient versus the temperature of cubic $PbREO_3\; (RE=Ce,\;Nd,\;Gd)$ perovskite oxides.} \label{Figure:7}
\end{figure}

To describe the ability to generate electric potential from a temperature gradient, we have calculated the Seebeck coefficient with respect to temperature as illustrated in figure \ref{Figure:7}. High Seebeck parameter is usually correspond on the good thermoelectric material \cite{41}. 
According to the spectra, it can be seen that Seebeck coefficient increases with increasing temperature. 
For all perovskites, the increasing temperature increases the energy per charge and eventually increases the voltage. The $PbGdO_3$ and $PbNdO_3$ exhibit positive values of $S$, indicating that the majority of carriers are holes. Whereas, $PbCeO_3$ presents a negative value which shows that the majority of carriers are electrons.\\
At room temperature, The calculated values of the Seebeck coefficient are mentioned in Table \ref{Table:6}. In our computations, at room temperature, the maximum value noted of the Seebeck coefficient was for $PbNdO_3$ which equal to $74.4\; \mu V.K^{-1}$. The Seebeck coefficient at $200K$ for $PbGdO_3$ was $30.5\;\mu V.K^{-1}$. 
As the temperature rose, the seebeck constant of $PbGdO_3$ increases linearly which is equal to $40.6\;\mu V.K^{-1}$ at $300K$ and finally with $105\; \mu V.K^{-1}$ at $900K$. In the case of $PbCeO_3$, the curve of $S$ increases with temperature variation and in room temperature ($300K$), $S$ is found to be $-41.8\;\mu V.K^{-1}$. As the temperature increases from $300K$, the $S$ value increases up to $-14.7\;\mu V.K^{-1}$ at $900K$.
\begin{table}[h]
	\begin{center}
		\begin{tabularx}{1.0\textwidth} { 
				>{\raggedright\arraybackslash}X 
				>{\centering\arraybackslash}X
				>{\centering\arraybackslash}X      
				>{\centering\arraybackslash}X    
				>{\raggedleft\arraybackslash}X}
			\hline	
			Compounds                   & $S\;(\mu V/K)$  & $\sigma/\tau\;(1/\Omega.m.s)$ & $PF\;(W/m.s.K^{2})$  \\
			\hline     
			$PbCeO_3$ \vspace{0.05cm}     & $-41.8$           & $2.79\;10^{20}$                         & $5.65\;10^{11}$   \\
			$PbNdO_3$ \vspace{0.05cm}     & $74.4$            & $3.09\;10^{20}$                         & $1.13\;10^{11}$   \\	
			$PbGdO_3$ \vspace{0.05cm}     & $40.6$            & $2.59\;10^{20}$                         & $1.04\;10^{11}$   \\
			\hline
		\end{tabularx}
		\caption {Calculated values of the thermoelectric parameters at room temperature of $PbREO_3\; (RE=Ce,\;Nd,\;Gd)$ compounds.}
		\label{Table:6}
	\end{center}
\end{table}
\subsubsection*{Electrical conductivity:}
\begin{figure}[H]
	\centering
	\includegraphics[scale=0.65]{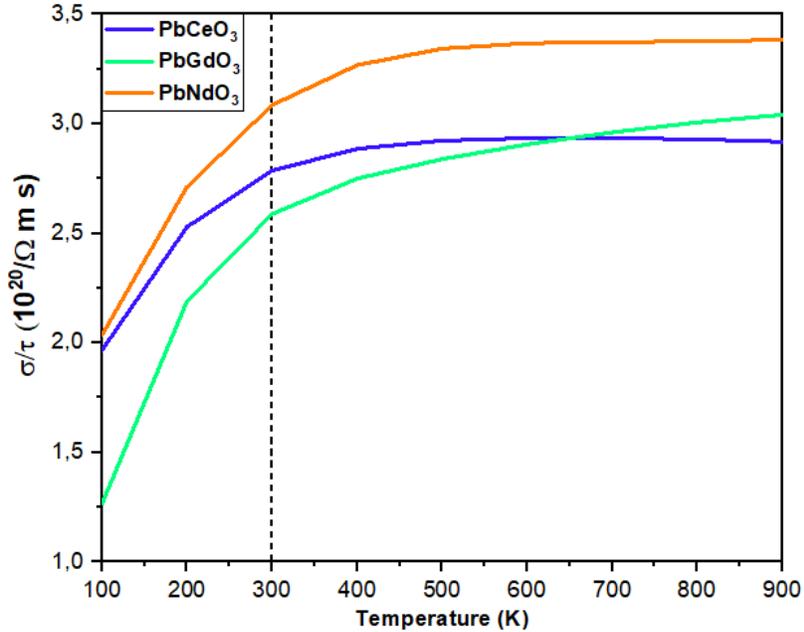}
	\caption{Electrical conductivity versus the temperature of cubic $PbREO_3\; (RE=Ce,\;Nd,\;Gd)$ perovskite oxides.} \label{Figure:8}
\end{figure}

The electrical conductivity per relaxation time $\sigma / \tau$ of all our compounds is investigated as illustrated in Figure \ref{Figure:8}. It is defined as the movement of the electrons through a material. For a semiconductor nature, the holes and electrons are responsible for the electrical conductivity.\\
From figure \ref{Figure:8}, we notice that $\sigma / \tau$ of these perovskite oxides is of the order $10^{20}$ $(\Omega m s)^{-1}$. 
It has been seen that the electrical conductivity per relaxation time ($\sigma / \tau$) increases with temperature, and this indicates the semiconductor nature of these compounds with conductivity values of $1.97\times10^{20}(\Omega m s)^{-1}$, $0.16\times10^{20}(\Omega m s)^{-1}$ and $2.04\times10^{21}(\Omega m s)^{-1}$ respectively for $PbCeO_3$, $PbGdO_3$ and $PbNdO_3$. The calculated values of $\sigma / \tau$ at room temperature are tabulated in table \ref{Table:6}. $PbNdO_3$ has a maximum value of the $\sigma / \tau$.
\subsubsection*{Power Factor:}
\begin{figure}[H]
	\centering
	\includegraphics[scale=0.7]{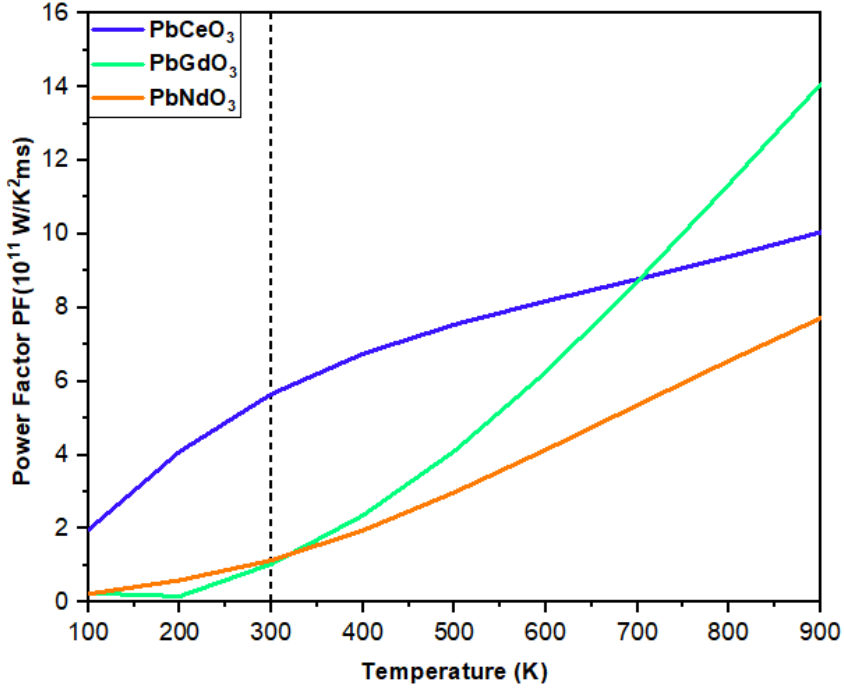}
	\caption{Power factor with respect to temperature of cubic $PbREO_3\; (RE=Ce,\;Nd,\;Gd)$ perovskite oxides.} \label{Figure:9}
\end{figure}

To show the thermoelectric efficiency of each material, we have calculated the power factor ($PF$) as a function of temperature for cubic $ PbREO_{3}\; (RE=Ce,\; Nd,\; Gd)$ as illustrated in figure \ref{Figure:9}. It has been proved that increasing temperature, the $PF$ increases. We notice that $PbGdO_3$ has a maximum power factor of $1.41\;10^{12}W.K^{2}.m.s$ at $900K$. At room temperature, the calculated values of the electronic power factor are tabulated in table \ref{Table:6}. We note that $PbCeO_3$ has a higher $PF$ than others about $5.65\;10^{11}W.K^{2}.m.s$ which can be considered as an efficient TE material.
\newpage
\section{Conclusion}
First-principle study using the FP-LAPW method is used to predict the structural, elastic, electronic, optical and thermoelectric properties of $PbCeO_3$, $PbNdO_3$ and $PbGdO_3$ compounds. The structural and elastic properties present a stable behavior of the studied compounds. Furthermore, the electronic properties indicate the semiconductor behavior of all studied materials with direct band gaps which are equal to $2.60\;eV$, $1.02\;eV$ and $1.55\;eV$ respectively for $PbCeO_3$, $PbNdO_3$ and $PbGdO_3$. 
The investigation of the optical properties such as dielectric function, refraction index, reflectivity and absorption coefficient of these materials present an important absorption coefficient in the visible region. The thermoelectric properties show that the majority of charge carriers responsible for the conduction of the $PbNdO_3$ and $PbGdO_3$ are holes while they are electrons for $PbCeO_3$. They also present high electrical conductivity with a good power factor, which indicates a desirable effect in TE materials.As a result, all these properties of cubic perovskite oxides $PbREO_3\; (RE=Ce,\; Gd,\; Nd)$ give a new way for efficient applications in thermoelectric and optoelectronic.

\end{document}